\documentclass[preprint,showpacs,preprintnumbers,nobibnotes]{revtex4}
\usepackage{amssymb}
\usepackage{amsmath}
\usepackage{graphicx}
\usepackage{dcolumn}
\usepackage{bm}
\usepackage{epsfig}

\input{tcilatex}
\begin{document}

\preprint{}
\title{Long-time tails in the random transverse Ising chain}
\author{Zhong-Qiang Liu$^{1\text{,}2}$}
\author{Xiang-Mu Kong$^{2}$}
\altaffiliation{Corresponding author. E-mail address: kongxm@mail.qfnu.edu.cn}
\author{Su-Rong Jiang$^{2\text{,}3}$}
\author{Ying-Jun Li$^{1}$}
\altaffiliation{Corresponding author. E-mail address: lyj@aphy.iphy.ac.cn}
\affiliation{$^{1}$State Key Laboratory for GeoMechanics and Deep Underground
Engineering, SMCE, China University of Mining and Technology, Beijing
100083, China \\
$^{2}$Shandong Provincial Key Laboratory of Laser Polarization and
Information Technology, Department of Physics, Qufu Normal University, Qufu
273165, China\\
$^{3}$Qindao College, Qingdao Technological University, Qingdao 266106, China}
\date{\today }

\begin{abstract}
Taking one-dimensional random transverse Ising model (RTIM) with the
double-Gaussian disorder for example, we investigated the spin
autocorrelation function (SAF) and associated spectral density at high
temperature by the recursion method. Based on the first twelve recurrants
obtained analytically, we have found strong numerical evidence for the
long-time tail in the SAF of a single spin. Numerical results indicate that
when the standard deviation $\sigma_{JS}$ (or $\sigma_{BS}$) of the exchange
couplings $J_{i}$ (or the random transverse fields $B_{i}$) is small, no
long-time tail appears in the SAF. The spin system undergoes a crossover
from a central-peak behavior to a collective-mode behavior, which is the
dynamical characteristics of RTIM with the bimodal disorder. However, when
the standard deviation is large enough, the system exhibits similar dynamics
behaviors to those of the RTIM with the Gaussian disorder, i.e., the system
exhibits an enhanced central-peak behavior for large $\sigma_{JS}$ or a
disordered behavior for large $\sigma_{BS}$. In this instance, the long-time
tails in the SAFs appear, i.e., $C(t)\sim t^{-2}$. Similar properties are
obtained when the random variables ($J_{i}$ or $B_{i}$) satisfy other
distributions such as the double-exponential distribution and the
double-uniform distribution.
\end{abstract}

\maketitle

The time-dependent behavior of quantum spin systems has been the subject of
theoretical studies for quite some time. However, most literature deals with
the dynamic correlation functions of the pure quantum spin systems. It is
found that the spin autocorrelation functions (SAFs) of these pure systems
are associated with the temperature. The transverse correlation functions
show a power-law behavior at $T=0$\cite{H.G. Vaidya788384}, an exponential
behavior at $0<T<\infty $\cite{A. R. Its9395}, and a Gaussian behavior at $%
T=\infty $\cite{A. Sur757677,J. F. Lee1987}, respectively. As to the
longitudinal correlation functions, Niemeijer\cite{Niemeijer1967} found that
SAF of the XY spin chain is a damped oscillation function which may be
expressed by the square of the Bessel function at $T=\infty$, but it has a
more complex expression for other temperature\cite{S. Katsura1970}. For the
spin-1/2 van der Waals model, Dekeyser and Lee\cite{Dekeyser1991} pointed
out that the power-law long-time tails are found in the transverse component
only at high temperature. Only recently, since the disorder effects besides
the effects of the temperature has been shown to affect the dynamics
behavior of spin systems in a drastic way providing a very rich area of
investigation, more and more attention has been paid on the random quantum
spin systems, especially on the low-dimensional systems, such as the random
transverse Ising model (RTIM)\cite{H. Rieger1997,A. P. Young1997,J. F1999,B.
Boechat2000,Zhq2006,Chen2010,XJia2006}, XY chain\cite{M. E.2003,M.
E.2004,Xu2008}, XX and XXZ chain\cite{Nicolas2004}, etc.. The known results
indicate that the disorder effects may drastically slow down the behavior of
the disordered averaged spin-spin correlation function\cite{Nicolas2004},
including the SAFs. For example, as the exchange couplings vary, the
Gaussian behavior of $xx$-relaxation of the single impurity in XY chain
slows down to the stretched exponential-like relaxation at $T=\infty $\cite%
{S. Sen1998}. However, as far as we know, although the short-time behaviors
of the SAFs have been reported\cite{J. F1999,B. Boechat2000,Zhq2006,M.
E.2003,M. E.2004,Xu2008,Chen2010}, the long-time behaviors of them are still
not understood. An important issue in this work is to explore whether it is
possible to realize a situation in which the exchange couplings and the
transverse fields are independent random variables, one can obtain the
long-time behavior of SAF of the random transverse Ising chain.

One-dimensional (1-D) RTIM is one of simplest but nontrivial example of the
random quantum spin chains. It is defined by the following Hamiltonian:%
\begin{equation}
H=-1/2\sum_{i}J_{i}\sigma _{i}^{x}\sigma _{i+1}^{x}-1/2\sum_{i}B_{i}\sigma
_{i}^{z}\text{,}  \label{hami}
\end{equation}%
where $1/2$ is the constant number introduced for the convenience of the
derivation. $\sigma _{i}^{x}$, $\sigma _{i}^{z}$ denote Pauli matrices at
site $i$. The exchange couplings $J_{i}$\ (or the transverse fields $B_{i}$)
are independent random variables obeying the distribution $\rho (J_{i})$ (or 
$\rho (B_{i})$). This model has a well-known physical interpretation in
connection not only with some quasi-one-dimensional hydrogen-bonded
ferroelectric crystals like Cs(H$_{1-x}$D$_{x}$)$_{2}$PO$_{4} $, PbH$_{1-x}$D%
$_{x}$PO$_{4}$ etc.\cite{J. A.1982,S. Watarai8485}, but also with Ising spin
glasses LiHo$_{0.167}$Y$_{0.833}$F$_{4}$\cite{W. Wu1991}. Besides, this
model may capture vital features of the recent neutron scattering experiment
in LiHoF$_{4}$\cite{XJia2006,H. M.2005}.

In this rapid communication, we investigate the dynamics of RTIM with the
double-Gaussian disorder, i.e., $J_{i}$\ or $B_{i}$ respectively satisfy the
double-Gaussian distribution 
\begin{equation}
\rho (\beta _{i})=p\rho _{1}(\beta _{i})+(1-p)\rho _{2}(\beta _{i}),
\label{Jdist}
\end{equation}%
where%
\begin{equation}
\rho _{1}(\beta _{i})=\frac{1}{\sqrt{2\pi }\sigma _{\beta }}\text{exp}\left[
-(\beta _{i}-\beta _{1m})^{2}/2\sigma _{\beta }^{2}\right]  \label{Gaussian1}
\end{equation}%
and%
\begin{equation}
\rho _{2}(\beta _{i})=\frac{1}{\sqrt{2\pi }\sigma _{\beta }}\text{exp}\left[
-(\beta _{i}-\beta _{2m})^{2}/2\sigma _{\beta }^{2}\right]  \label{Gaussian2}
\end{equation}%
are the standard Gaussian distributions, wherein $\beta _{1m}$ ($J_{1m}$ or $%
B_{1m}$) and $\beta _{2m}$ ($J_{2m}$ or $B_{2m}$) denote the corresponding
mean values of the random variables $\beta _{i}$ ($J_{i}$ or $B_{i}$), $%
\sigma _{\beta }$ ($\sigma _{J}$ or $\sigma _{B}$) is the standard
deviation. The physical sense of Eq. (\ref{Jdist}) is that the random
variable $\beta _{i}$ satisfies $\rho _{1}(\beta _{i})$ with probability $p$
and $\rho _{2}(\beta _{i})$ with probability $(1-p)$, respectively. The mean
value and the standard deviation of the random variables $\beta _{i}$
obeying distribution $\rho (\beta _{i})$ are respectively $p\beta
_{1m}+(1-p)\beta _{2m}$ and $\sigma _{\beta S }=\sqrt{p(1-p)(\beta
_{1m}-\beta _{2m})^{2}+\sigma _{\beta }^{2} }$. Since $\sigma _{\beta S }$
is an increasing function of $\sigma _{\beta}$, for convenience, $\sigma
_{\beta}$ is used to denote $\sigma _{\beta S }$ in the following unless
specified otherwise.

Obviously, the cases discussed in references \cite{J. F1999} and \cite%
{Zhq2006} are two special cases of the RTIM with the double-Gaussian
disorder. When $\sigma _{\beta }\rightarrow 0$, Eq. (\ref{Jdist}) becomes
the bimodal distribution%
\begin{equation}
\rho (\beta _{i})=p\text{ }\delta (\beta _{i}-\beta _{1m})+(1-p)\text{ }%
\delta (\beta _{i}-\beta _{2m}).  \label{bimodal}
\end{equation}%
When $p=1$ (or $p=0$), Eq. (\ref{Jdist}) becomes a standard Gaussian
distribution. Of course, $\rho _{1}(\beta _{i})$ and $\rho _{2}(\beta _{i})$
may be other distributions, such as an exponential distribution or a uniform
distribution. In this way, $\rho (\beta _{i})$ becomes the
double-exponential distribution or the double-uniform distribution.

The basic tool employing in this work is the recursion method\cite{V. S.
Viswanath1994}, which is very powerful in the study on many-body dynamics%
\cite{J. F1985,M.H. Lee828365,J. F1999,Zhq2006,Xu2008,M. E.2003,M. E.2004,J.
F. Lee1987,S. Sen9195,J.O.F.1997,J. Hong19951997}. It may help us obtain the
SAF and associated spectral density. They are respectively defined as%
\begin{equation}
C\left( t\right) =\overline{\left\langle \sigma _{j}^{x}\left( t\right)
\sigma _{j}^{x}\left( 0\right)\right\rangle },  \label{C(t)}
\end{equation}%
and%
\begin{equation}
\Phi \left( \omega \right) =\int_{0}^{+\infty }dtC\left( t\right)
e^{-i\omega t},  \label{Ftransform}
\end{equation}
where $\overline{\left\langle \cdots \right\rangle}$ denotes an average over
the random variables is performed after the statistical average. At $%
T\rightarrow\infty$, the SAF may be expanded into a power series 
\begin{equation}
C\left( t\right) =\sum_{k=0}^{\infty }\frac{\left( -1\right) ^{k}}{\left(
2k\right) !}\mu _{2k}t^{2k},  \label{C(t)expansion}
\end{equation}%
where $\mu _{2k}\ $is the $2k$th moment of the SAF. And the spectral density
may be determined via the relation%
\begin{equation}
\Phi \left( \omega \right) =\lim_{\varepsilon \rightarrow 0}Rea_{0}(z)\mid
_{z=\varepsilon +i\omega },  \label{SpDLT}
\end{equation}
where%
\begin{equation}
a_{0}(z)=1/\left( z+\Delta _{1}/\left\{ z+\Delta _{2}/\left[ z+\Delta
_{3}/\left( z+...\right) \right] \right\} \right) ,  \label{CtCfraction}
\end{equation}
which is the Laplace transform of the SAF (\ref{C(t)}). The
continued-fraction coefficients (recurrants) $\Delta _{\nu }$ ( $\nu=1, 2,
\cdots$) of the SAF may be derived by the recurrence relation I given in
reference\cite{V. S. Viswanath1994}.

Although only a finite number of recurrants can be deduced analytically,
they play vital roles in the whole study process. On the one hand, since the
first $2k$ moments may be expressed by the first $k$ recurrants\cite{V. S.
Viswanath1994}, SAF can be determined by the Pad\'{e} approximant
constructed out of the known moments. On the other hand, one can determine
the spectral density of the SAF by employing the known recurrants and
so-called Gaussian terminator\cite{V. S. Viswanath1994,J.O.F.1997,J.
Stolze1992,J. F1999}. Due to the first $12$ recurrants have been deduced
analytically, we extract more relevant information from them in this paper
than we did in our previous work\cite{Zhq2006} in which only the first $9$
recurrants have been obtained analytically.

In this work we analyze two types of RTIM: a random-bonds model, and a
random-fields model. In the former, the exchange couplings $J_{i}$\ are
chosen independently from the double-Gaussian distributions defined by Eq. (%
\ref{Jdist}), while the transverse fields $B_{i}$ remain unaltered. To
compare our results with the previous results presented in reference \cite%
{J. F1999}, we set the same parameters as those in it: $J_{1m}=1.0$, $%
J_{2m}=0.4$ and $B_{i}=B=1$\ which fixes the energy scale, and $\sigma _{J}$
varies from $0.01$ to $2$. $C\left( t\right) $ and $\Phi \left( \omega
\right)$ are plotted in Fig. 1 and Fig. 2, respectively. Each inset in Fig.
2 gives the recurrants $\Delta _{\nu }$ $(\nu =1,\cdots ,12)$ for the same
parameters in each figure, and the lines in it are just a guide to the eye.

As $\sigma _{J}\rightarrow 0$, the double-Gaussian distribution turns into
the bimodal distribution. Taking $\sigma _{J}=0.01$ for example, as shown in
Figs. 1 (a) and 2 (a), the dynamics of the system undergoes a crossover from
a central-peak behavior to a collective-mode behavior as $p$ decreases from $%
1$ to $0$. Although we can not obtain $C\left( t\right)$ for large time,
comparison shows that Fig.1 (a) offers a better improvement in SAFs for
short time than those given by Florencio, et al.\cite{J. F1999}. Comparing
all the figures in Fig. 1 and Fig. 2, one finds that as $\sigma _{J}$
increases from $0.01$ to $2$, the damped oscillations of the SAFs diminish
in magnitude, meanwhile the recurrants for the same parameters tend to fall
into a straight line. In this process, the spectral weights at $\omega=0$
are enhanced, and the peaks' locations of the spectral lines move towards $%
\omega=0$ gradually (e.g., line $p=0$). The solid lines with $p=1$ in Fig. 2
illustrate how an almost pure Gaussian profile of $\Phi (\omega )$ evolves
into a curve with some additional structure which consists of a central peak
of increasing height and decreasing width. Above analysis indicates that the
collective-mode behavior becomes weaker and weaker, while the dynamics
becomes increasingly dominated by a central-peak behavior as $\sigma _{J}$
increases. When\ $\sigma _{J}$\ are large enough (see Figs. 1(d) and 2(d)),
the crossover vanishes, and the system exhibits an enhanced central-peak
behavior. These results are in agreement with those in our previous work\cite%
{Zhq2006}.

The infrared (i.e., low-frequency) singularity in the spectral density
signals the SAF should manifest itself a characteristic quantum-diffusion
long-time tail\cite{M. B1994}. We find, as expected, long-time tails exist
in the SAFs when $\sigma _{J}$ are large enough. Fig. 3 gives the log-log
plot of the absolute values of $C\left( t\right) $ versus time. The solid
lines with $p=1$ in Figs. 1 and 3 show how a Gaussian behavior of $C\left(
t\right)$, due to the disorder effects, slows down to a power-law
relaxation. Calculations indicate that $C\left( t\right) \sim t^{-2}$ for $%
t>20$. This result is identical with the pure XY interaction case ($R=0$) of
the spin-$1/2$ van der Waals model\cite{Dekeyser1991}. In fact, the XY model
and the transverse Ising model are dynamically equivalent to the spin van
der Waals model\cite{J. F. Lee1987}. These systems are indicated by the same
geometry of their respective dynamics Hilbert spaces. Finally, we should
like to emphasize that success in obtaining the long-time tails of the SAFs
is owing to the fact that the true long-time behavior is only nebulously
encoded in the first few continued-fraction coefficients\cite{M. B1994}.

For the random-fields model, the random transverse fields $B_{i}$ are
independent and satisfy the double-Gaussian distributions given by Eq. (\ref%
{Jdist}), while the exchange couplings are uniform (i.e., $J_{i}=J=1$). For
convenience to compare our results with the known results\cite{J. F1999}, we
set $B_{1m}=0.6$, $B_{2m}=1.4$, and $\sigma _{B}$ varies from $0.01$ to $2 $%
. The SAFs and the corresponding spectral densities are shown in Fig. 4 and
Fig. 5, respectively. The inserts in them respectively give the
double-logarithmic plots of the absolute values of $C\left( t\right) $
versus time and the associated recurrants $\Delta _{\nu }$ $(\nu =1,\cdots
,12)$ for the same parameters in each figure.

When $\sigma _{B}$ is small (e.g., $\sigma _{B}\leq 0.10$), the system
undergoes a crossover from a central-peak behavior to a collective-mode
behavior as $p$ decreases from $1$ to $0$, as shown in Figs. 4 (a) and 5
(a). This result is consistent with that obtained by Florencio, et al.\cite%
{J. F1999}. However, as $\sigma _{B}$ increases, this crossover disappears
gradually, see Fig. 5 (b), (c) and (d). Meanwhile, the SAFs exhibit an
interesting phenomenon. From Fig. 4 (b) we find that the decays of long-time
tails in lines $p=0$ and $p=1$ are proportional to $-t^{-2}$, and those in
other lines are proportional to $t^{-2}$. When $\sigma _{B}$ is large enough
(e.g., $\sigma _{B}= 1.00$), SAFs fall into three successive phases. In the
initial phase ($0<t<1$), the decay of $C\left( t\right) $ is Gaussian
regime. In the third phase ($t>20$), the long-time tails of SAFs obey the
power-law decay $t^{-2}$. There is a cross-over time between the Gaussian
regime and the power-law regime. In this cross-over, the oscillation of $%
C\left( t\right) $ implies that the $x$ component of the spin may be
coherently coupled to some mode with small fluctuation\cite{Dekeyser1991}.
Thus, slow decay may emerge. The strong peaks at $\omega=0$ aslo signal the
existence of the long-time tails in the SAFs. The cross-over time becomes
shorter as $\sigma _{B}$ is increased further. While more and more
high-frequency modes are involved in the corresponding spectral densities.
The system is in a disordered state when $\sigma _{B}$ is large enough.

Above we have investigated the dynamics of the RTIM with the double-Gaussian
disorder at high temperature by the recursion method. This generalized model
may recover the dynamics of both the RTIM with the bimodal disorder and that
with the Gaussian disorder. One may wonder if these spin dynamics features
are universal or not. Therefor, we have studied the dynamics of the RTIM
with other types of disorder, in which the variables independently satisfy
the double-Exponential distribution, and the double-Uniform distribution.
Numerical results indicate that these systems show a similar dynamical
behavior to those of the RTIM with the double-Gaussian disorder. Generally
speaking, when $\sigma _{JS}$ (or $\sigma _{BS}$) is small, the dynamics of
the system undergoes a crossover from a central-peak behavior to a
collective-mode one as $p$ (or the mean value of $J_{i}$\ or $B_{i}$ )
varies. No long-time tail in SAF is obtained. As $\sigma _{JS}$ (or $\sigma
_{BS}$) increases, the crossover vanishes gradually, and the power-law decay
appears in the SAFs when the standard deviation is large enough. For the
case of large $\sigma _{JS}$ the system exhibits a central-peak behavior,
but it shows a disordered behavior for the case of large $\sigma _{BS}$. As
expected, we found that, independently of the type of disorder, the SAFs in
1-D RTIM decay as $t^{-2}$ for large time due to quantum spin diffusion.

This work is supported by National Basic Research Program of China under
Grant No. 2007CB815105, National Natural Science Foundation of China under
Grant No. 11074300, 10775088, and the Specialized Research Fund for the
Doctoral Program of Higher Education under Grant No. 2010370511000.

\begin{center}
\bigskip \textbf{Figure Captions}
\end{center}

Fig. 1. (Color online) Spin autocorrelation functions $C\left( t\right) $ of
RTIM with the double-Gaussian random bonds. Better improvement in SAFs is
made than those given in references \cite{J. F1999} and \cite{Zhq2006}.

\bigskip 

Fig. 2. (Color online) The associated spectral densities for the same
parameters as in Fig. 1. The insets show the recurrants $\Delta _{\nu }$ $%
(\nu =1,\cdots ,12)$. The lines in the insets are just a guide to the eye.
For small\ $\sigma _{J}$ (e.g., $\sigma _{J}\leq 0.01$), (a) recovers $\Phi
(\omega )$ of RTIM with the bimodal distribution. As $sigma_{J}$ increases,
the crossover from a central-peak behavior to a collective-mode behavior
vanishes gradually. The system only exhibits an enhanced central-peak
behavior when $\sigma _{J}$\ is large enough.

\bigskip 

Fig. 3. (Color online) Log-log plot of $\left\vert C\left( t\right)
\right\vert $ versus $t$ for different values of $p$ and $\sigma _{J}$.
Black dashed line in each figure represents a power-law decay $t^{-2}$.

\bigskip 

Fig. 4. (Color online) SAFs $C\left( t\right) $ of RTIM with the
double-Gaussian random fields. When $\sigma _{B}$ is small (e.g., $\sigma
_{B}=0.01$), (a) approximately recovers the results presented in Ref. \cite%
{J. F1999}. The inserts present log-log plot of $\left\vert C\left( t\right)
\right\vert $ versus $t$. The black dashed line in each insert represents a
power-law decay $t^{-2}$.

\bigskip 

Fig. 5. (Color online) The corresponding spectral densities for the same
parameters as in Fig. 4. For small $\sigma _{B}$ (e.g., $\sigma _{B}=0.01$),
the system moves from a collective mode dynamics to a central peak type of
dynamics as $p$ varies from $0$ to $1$. However, the system is in a
disordered state when $\sigma _{B}$\ is large enough (e.g., $\sigma
_{B}=2.00 $).

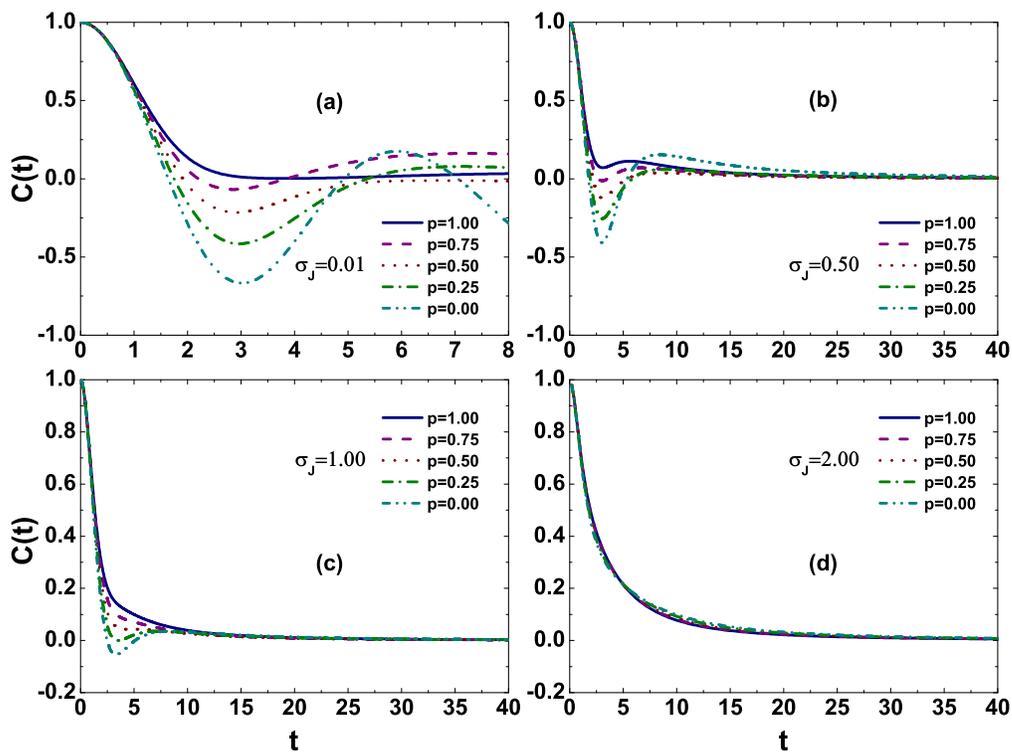
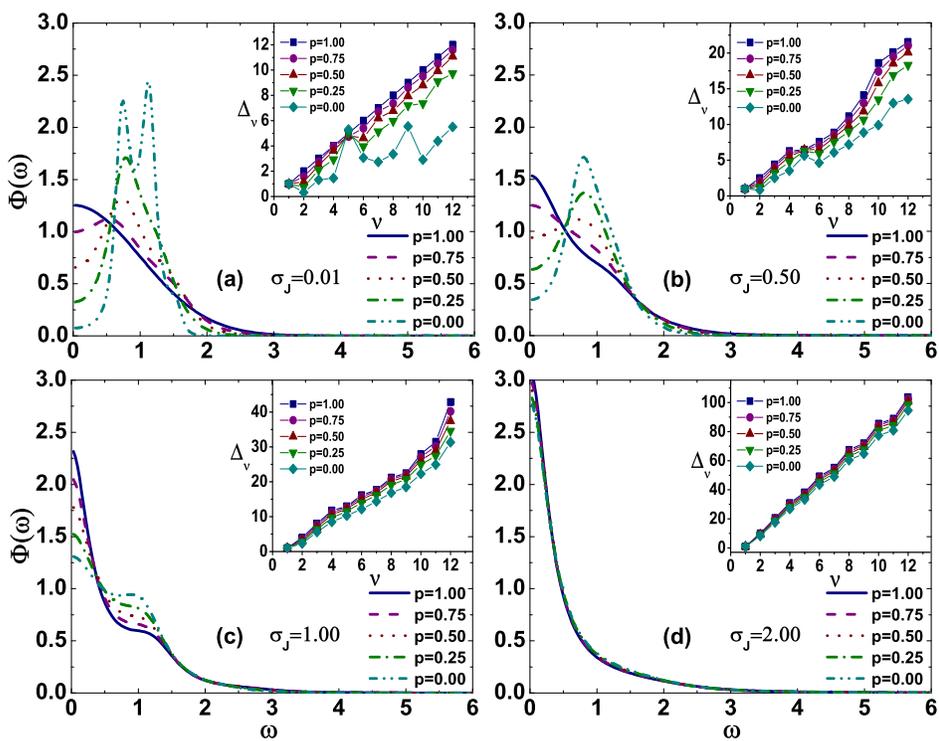
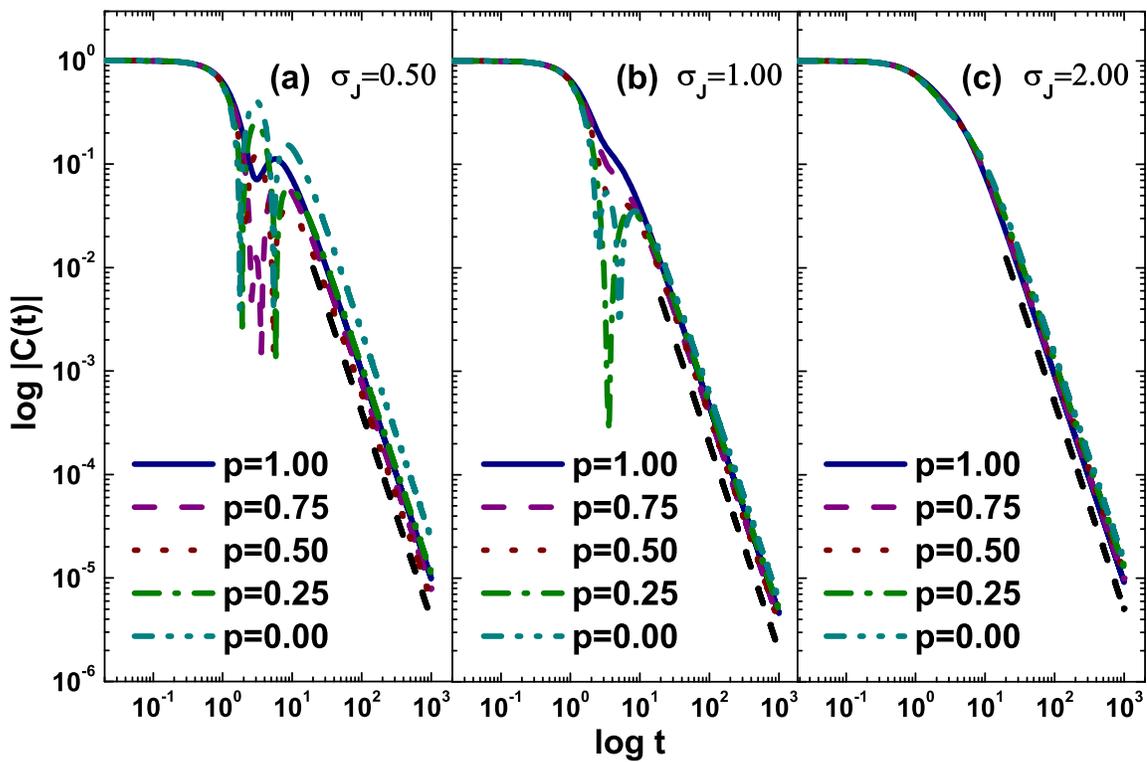
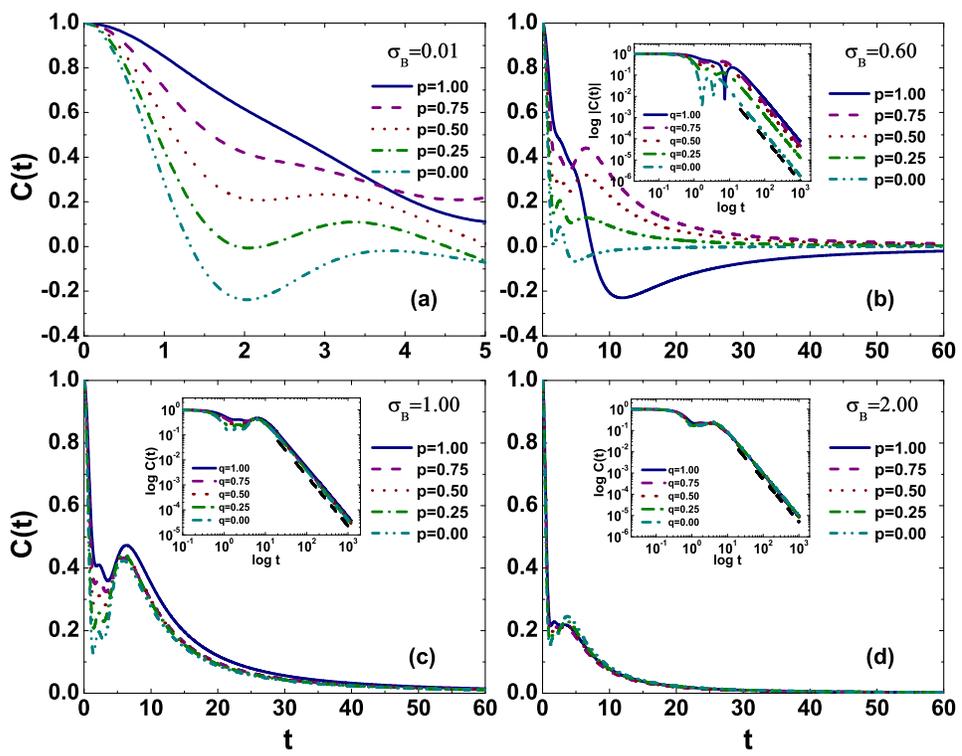
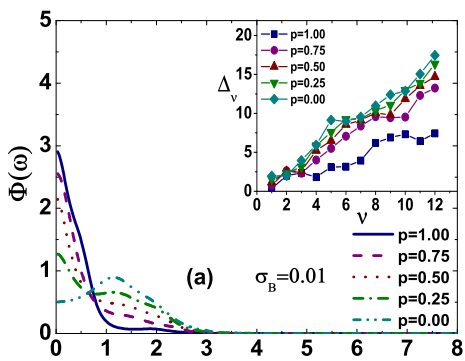
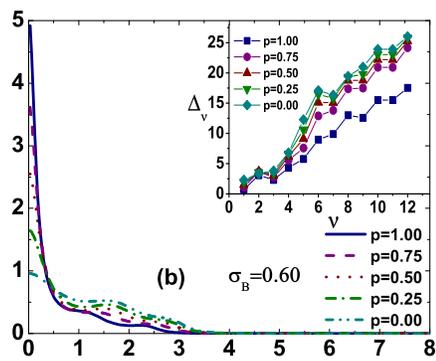
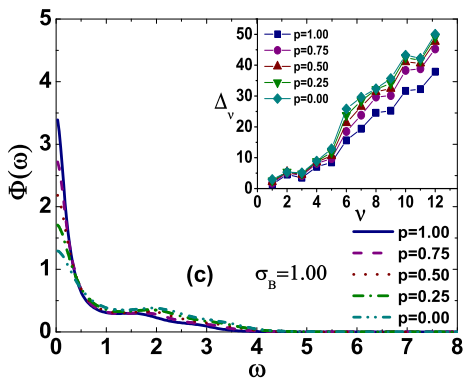
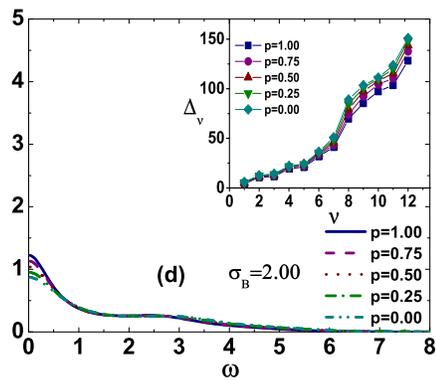
\end{document}